# Temperature-induced suppression of structural disproportionation in paramagnetic quantum materials

Himanshu Joshi[1], Mateusz Wlazło[1], Harshan Reddy Gopidi[1], Oleksandr I. Malyi[1,#]

[1]Centre of Excellence ENSEMBLE³ Sp. z o. o., Wolczynska Str. 133, 01-919, Warsaw, Poland

**#Email:** oleksandrmalyi@gmail.com (O.I.M)

**Abstract:**
With the development of electronic structure theory, a new class of materials - quantum ones has been recognized by the community. Traditionally, it has been believed that the properties of such compounds cannot be described within the framework of modern density functional theory, and indeed, more advanced post-mean-field theory methods are needed. Motivated by this, herein, we develop a fundamental understanding of such complex materials using the example of paramagnetic $YNiO_3$, which is experimentally known to exhibit metal-to-insulator phase transition. We show that this material has a temperature-dependent distribution of local structural and spin motifs. Thus, while at low temperatures, $YNiO_3$ has distinct structural disproportionation with the formation of large and small octahedra, as the temperature increases, this disproportionation is suppressed. We also explain the paramagnetic monoclinic to paramagnetic orthorhombic phase transition within the double-well to single-well energy profile, predicting the variation of the corresponding energy profile as a function of octahedral size distribution. In this way, we demonstrate a fundamental understanding of structural phase transitions in quantum materials, giving insight into how it can be used for different applications and what minimum level of theory is needed to describe such types of complex materials correctly.

**Introduction:**

Atomic identities, composition, and structure define the electronic properties of solid-state materials[1] and have been used to develop open-access databases (e.g., Materials Project[2], AFLOW[3], and OQMD[4]) containing electronic structures for thousands of materials. In recent years, it has also become clear that in addition to "classical" solid-state compounds (e.g., Si, ZnO), many quantum materials (especially paramagnetic phases)[5-15] exhibit a range of spin and/or structural symmetry breaking. For instance, in the case of magnetic materials, systems are considered to be magnetically long-range ordered until a certain critical temperature - the Néel temperature, above which the spin long-range order breaks, leading to the formation of a spin polymorphous system. Similarly, many quantum materials exhibit significant changes in local structural motifs with temperature - i.e., one associated with a monoclinic to orthorhombic or orthorhombic to cubic crystal structure phase transition[5]. The common tendency observed experimentally is that crystal structure symmetry increases with temperature[16], leading to a transition from multiple local environments to a single local environment.

Traditionally, the theoretical community has largely ignored the distribution of local motifs in quantum materials. This is primarily attributed to the deeply entrenched beliefs that the physics of these compounds can be explained through strong correlation[17-19], where the properties of materials are generally defined by the ratio of electron-electron repulsion (U) to bandwidth (W) without considering the distribution of local motifs. This perspective presumes that the interplay of electrons within a material – their correlations – dominates the behaviors we observe, such as conductivity, magnetism, and superconductivity. Recently, however, it has become evident that the role of strong correlation has been overestimated in many materials, and indeed, local motifs play the key role in determining material properties[5-15,20]. This also, in large part, challenges a correlation picture of bandwidth-driven insulator-to-metal phase transition (i.e., insulator-to-metal phase transition caused by reduction of U/W ratio) and highlights the possible role of the distribution of local motifs in materials properties. We note, however, that understanding of the distribution of local motifs and, more specifically, the temperature-dependent distribution of such motifs remains largely unexplored, especially in the strongly correlated compounds. For instance, in strongly correlated materials, the dynamic mean-filed theory has recently had a way to account for structural energy lowering[14,20-22] and even phonon calculation[23], but these methods are not yet routinely used for all studies. Moreover, for paramagnets, the theory of correlated materials is still usually based on density functional theory (DFT) for a global average structure where local and total magnetic moments are set to 0 (which is a naïve approximation of the paramagnets[5-7]) as the basis for more advanced theories.

Motivated by this, herein, we use the example of $YNiO_3$, which exists in three known experimentally-observed phases having different distributions of local motifs:

(i) The low-temperature α phase has a monoclinic crystal structure and an antiferromagnetic spin arrangement.[24] It exhibits structural disproportionation in the form of $NiO_6$ octahedral breathing mode (i.e., the defining feature of the monoclinic distortion) and magnetic disproportionation in the form of low and high spin sites.[9]

(ii) Above the Néel temperature of 145 K[24], YNiO$_3$ exists in the β phase, which retains the crystal structure and structural disproportionation of the α phase, but the magnetic long-range order is reduced to a paramagnetic arrangement with a net-zero magnetic moment. Importantly, it shows a distribution of local spins, which is crucial to maintain its insulating properties.[5-7]

(iii) At temperatures exceeding 582 K, the system undergoes a transition to an orthorhombic lattice system in the γ phase, concomitant with an insulator-to-metal transition[25], while both of the aforementioned disproportionation factors are washed out due to thermal motion.[5]

While the polymorphous theory of some representative quantum materials has been developed[5-15] and some fundamental physics of such compounds has been understood, the temperature-dependent physics of local motifs is rarely explored. Motivated by this, herein, we apply a spin polymorphous description of the paramagnetic phase within static and ab initio molecular dynamics (AIMD) to study the temperature-dependent distribution of local motifs in the β and γ phases. This approach allows us to observe the gradual softening of the structural disproportionation tendency, revealing the physics of phase transition and helping us to understand the temperature-induced changes in the distribution of local motifs and the origin of β and γ phase transitions.

**Methods:** All first-principles calculations were performed utilizing the Vienna An Initio Simulation (VASP) software package[26-30] with projector augmented-wave pseudopotentials[31] and SCAN functional[32]. The cutoff energy was set to 500 and 550 eV for final relaxation and volume optimization/molecular dynamics calculations, respectively. Volume optimization began with a 1000 k-point grid per reciprocal atom and was increased to 10,000 for the final internal structure relaxation. The AIMD calculations utilized a Γ point-only k-grid. The relaxation calculations were iteratively performed until the forces acting on each atom in the system were less than 0.01 eV/Å. The spin polymorphous model[13] of the PM phase with a 160-atom cell was used to describe the *β* phase, where spin distribution corresponds to a special quasi-random structure (SQS)[33] in contrast to the starting point in many other works which consider paramagnet as a global average structure with both local and total magnetic moments of 0. To investigate the impact of temperature on structural characteristics, AIMD using the NPT ensemble has been performed. In this ensemble, the number of atoms (N), pressure (P), and temperature (T) remain constant. The Langevin thermostat with Parrinello-Rahman barostat[34,35] was employed to maintain both temperature and pressure at consistent levels. The time step was set to 1 fs. The Langevin thermostat was utilized to establish a friction coefficient of 3 ps$^{-1}$ for the atomic degree of freedom of all atoms, while the lattice degree of freedom was set to 10 ps$^{-1}$. These calculations were performed with a cutoff energy of 550 eV and Γ-only k-grid. The data analysis and visualization were done using pymatgen[36] and Vesta software[37].

**Results and discussion:**

**Distribution of local motifs in paramagnetic YNiO$_3$ phases:** The paramagnetic β phase exhibits the structural and spin symmetry breaking, with the formation of two different octahedra resulting in the insulating state. This is well captured within the spin polymorphous theory[5-7] where the paramagnetic compound is considered within the spin SQS approximation[33]. This model aligns with the high-temperature limit of the paramagnetic phase[10,38], where the total magnetization is effectively zero, and each magnetic species exhibits an uncorrelated random spin distribution. Utilizing this spin SQS framework and the large supercell size needed to converge materials properties (see discussion in Ref. [39]), our SCAN calculations reveal a distribution of local $NiO_6$ octahedral volumes, with the resulting system having an insulating nature and band gap energy of 0.62 eV. Specifically, we observe the coexistence of large and small $NiO_6$ octahedra, measuring 10.61±0.05 Å$^3$ and 9.16±0.05 Å$^3$, respectively. Here, large and small octahedra are identified by the sublattice corresponding to the high (≈1.4 $\mu_B$) and low (≈0 $\mu_B$) spin states, respectively. We note that accounting for such distribution of local motifs is essential for describing the insulating nature of the β phase without accounting for any dynamic correlation effects, as discussed in our previous work.[7] We recognize the feasibility of considering intermediate-temperature spin configurations. Yet, the spin SQS model, which is commonly adopted in physics, seems adept at effectively representing the core physics of such systems[10]. It is also worth noting that in a spin polymorphous system, the orthorhombic $YNiO_3$ phase tends to disproportion unless the temperature is factored in.[5,40] This thus implies properties of orthorhombic $YNiO_3$ cannot be properly analyzed within the spin polymorphous theory that only relies on the minimization of internal energy. Indeed, this behavior is not surprising as at finite temperature, the behavior of a system is not defined anymore only by the internal energy minimization[41]. Hence, understanding of β and γ phase transition does require post-static calculations.

To examine the impact of temperature on the distribution of local structural motifs, we performed AIMD simulations using the NPT ensemble. It is important to mention that an increasing number of studies have performed similar simulations for different types of materials[10,42-49], but critical input parameters are not explicitly defined in the literature (for instance, the distribution of local motifs is often discussed for extremely short AIMD simulations with a total simulation time of a couple of picoseconds or even fewer). Moreover, $YNiO_3$ represents a unique set of compounds where both distributions of structural and spin motifs coexist simultaneously. For clarity, herein, our primary focus is to analyze structural motifs, which can be robustly studied experimentally using techniques like synchrotron x-ray diffraction(sXRD)[25] or local probe analysis of the pair distribution function[50,51]. While local probe measurements for spin motif distribution are feasible[52,53], they are less common and necessitate an advanced theoretical framework to account for various magnetic sublattices. Hence, the distribution of local spin motifs will not be discussed in detail here.

When given atomic identities, composition, and structure are introduced to AIMD, the atomic velocities are usually randomly assigned in accordance with the corresponding temperature. This thus initializes a system that is not explicitly adjusted for applied external conditions (e.g., temperature, pressure) and interatomic or intermolecular interactions. Because of this, there is a finite time for the system to reach equilibrium. Traditionally, such equilibration is verified by monitoring the time dependence of internal energy, temperature, or lattice parameters (Figs. 1-3). However, for quantum materials, the distribution of local motifs becomes pivotal, and hence achieving its convergence is critical. We should note that the time needed for convergence is influenced by the system proximity to the transition point and, indeed, may require sufficiently long simulation time and the use of post-AIMD methods (e.g., active learning MD[54-57]). Because of this, the exact temperature for the phase transition estimated from the AIMD simulation should be taken with caution, the most significant part that one can learn here is the temperature dependence of the distribution of local motifs and its impact on materials properties.

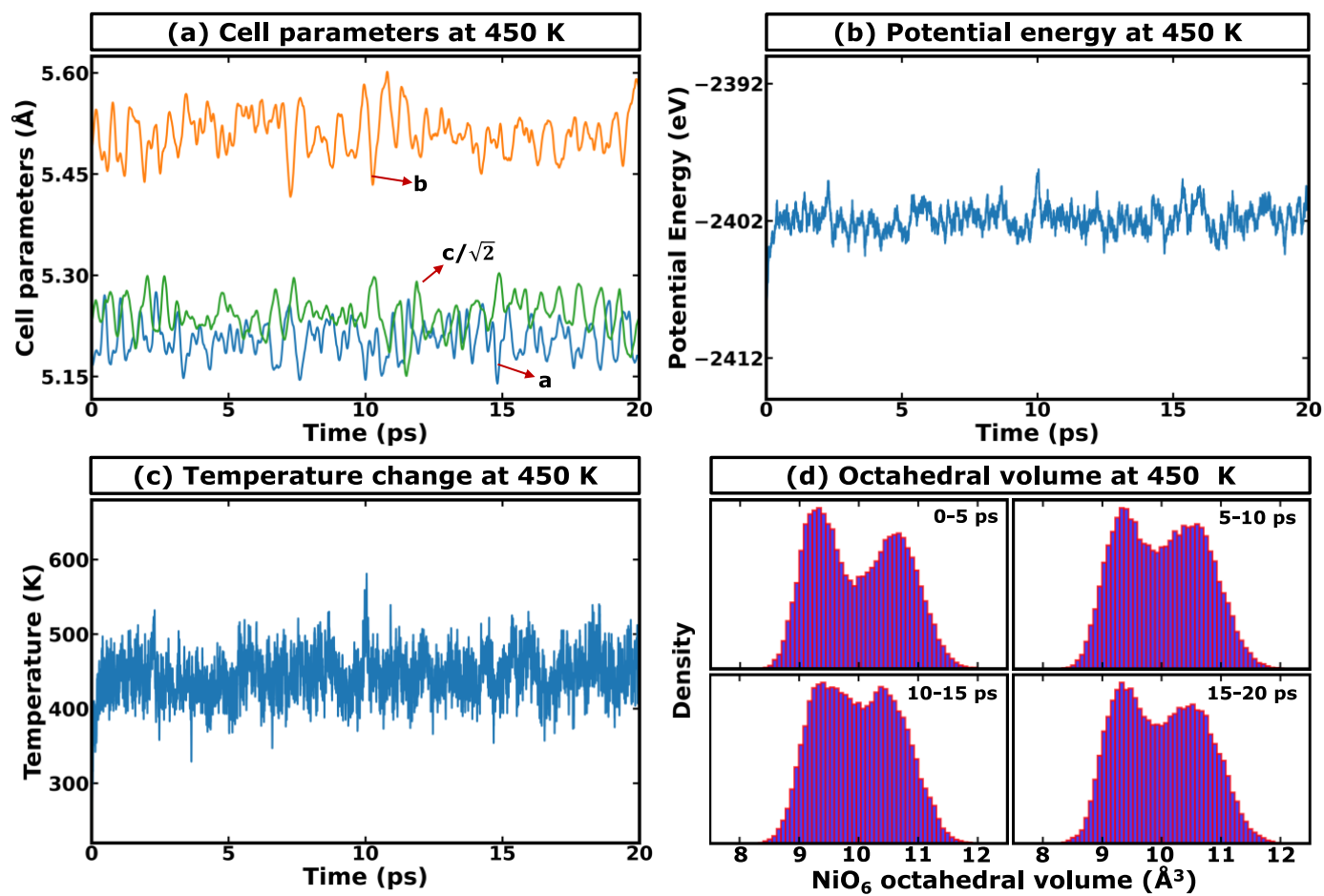

**Figure 1. First-principles molecular dynamics of PM phase of YNiO$_3$ at 450 K using a spin polymorphous model**. Time-dependent variation of (a) lattice constants, (b) potential energy, and (c) temperature during NPT molecular dynamics simulation at 450 K. (d) Distribution of local octahedral volumes at different moments of time.

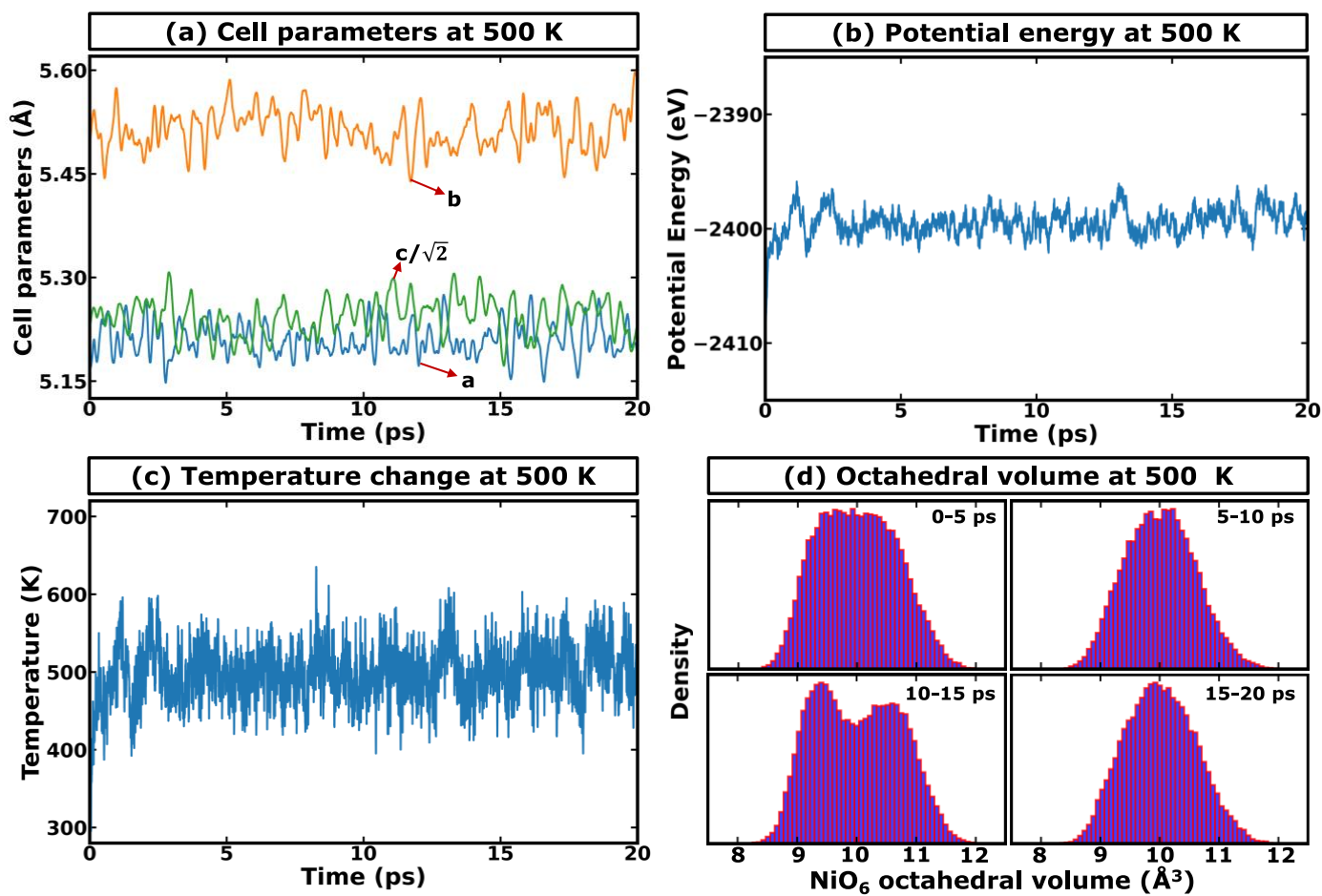

**Figure 2. First-principles molecular dynamics of PM phase of YNiO$_3$ at 500 K using a spin polymorphous model**. Time-dependent variation of (a) lattice constants, (b) potential energy, and (c) temperature during NPT molecular dynamics simulation at 500 K. (d) Distribution of local octahedral volumes at different moments of time.

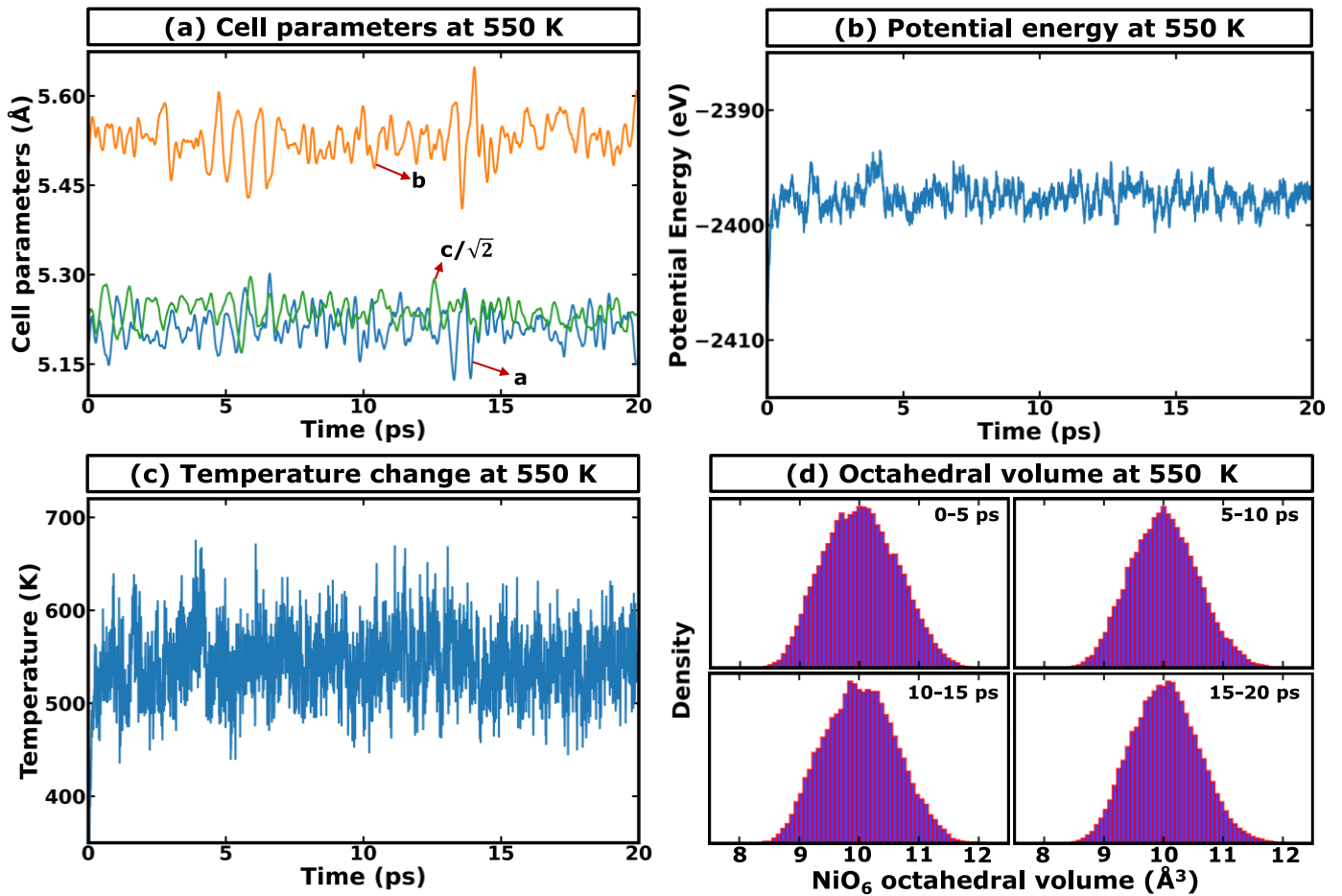

**Figure 3. First-principles molecular dynamics of PM phase of YNiO$_3$ at 550 K using a spin polymorphous model**. Time-dependent variation of (a) lattice constants, (b) potential energy, and (c) temperature during NPT molecular dynamics simulation at 550 K. (d) Distribution of local octahedral volumes at different moments of time.

**Temperature-dependent distribution of local motifs in paramagnetic YNiO$_3$ phases:** While at T=0 K calculations, the distribution of local motifs is not very significant. When the temperature is introduced, the intrinsic tendency of the system (defined by the minimization of internal energy) can be significantly modified by the temperature contribution (e.g., driven by entropy increase). This is indeed reflected in the distribution of local structural motifs, for simplicity calculated here as volumes of NiO$_6$ octahedra. Thus, at low temperatures, one can see bimodal Gaussian distribution around two distinct types of octahedral volumes (i.e., small and large). For instance, at 200 K, the average volumes of large and small octahedra are 10.67 and 9.19 Å$^3$, respectively. These results align with the static calculations for the β phase and indicate that at lower temperatures, the structure oscillates around the T=0 K spin polymorphous configuration. This observation underscores the effectiveness of the spin SQS model we developed. It also confirms that the T=0 K spin polymorphous configuration serves as a kernel for low-temperature AIMD simulations. As the temperature increases (Fig. 4), the two distinct octahedral peaks begin to shift slightly toward each other, and the distribution of local motifs becomes broader, with the two distributions starting to overlap. Eventually, at high temperatures (550 K), a broad-shifted Gaussian-like distribution of octahedral volumes emerges, and the average size of the octahedra is 10.03 Å$^3$. This shift in distribution, especially in higher temperature regimes, provides a deeper understanding of the phase transition. Indeed, the transition from double peak volume distribution to single peak volume distribution is responsible for the temperature-induced fall of the insulating state.[5] Taking into account that the monoclinic phase has a distinct disproportionation, it also becomes clear that the transition from bimodal to single Gaussian distribution is closely related to the monoclinic to orthorhombic structure phase transition, which is experimentally observed at approximately 582 K.[25] We emphasize that all these results are obtained with mean-field DFT without accounting for any additional dynamic correlation effect or using U correction, demonstrating the temperature-dependent suppression of the structural disproportionation. It is also important to note that while different Hamiltonian models usually consider the existence of only a single local environment, the current results clearly demonstrate that at all temperatures, there is a wide distribution of local structural environments, which, however, noticeably changes with temperature.

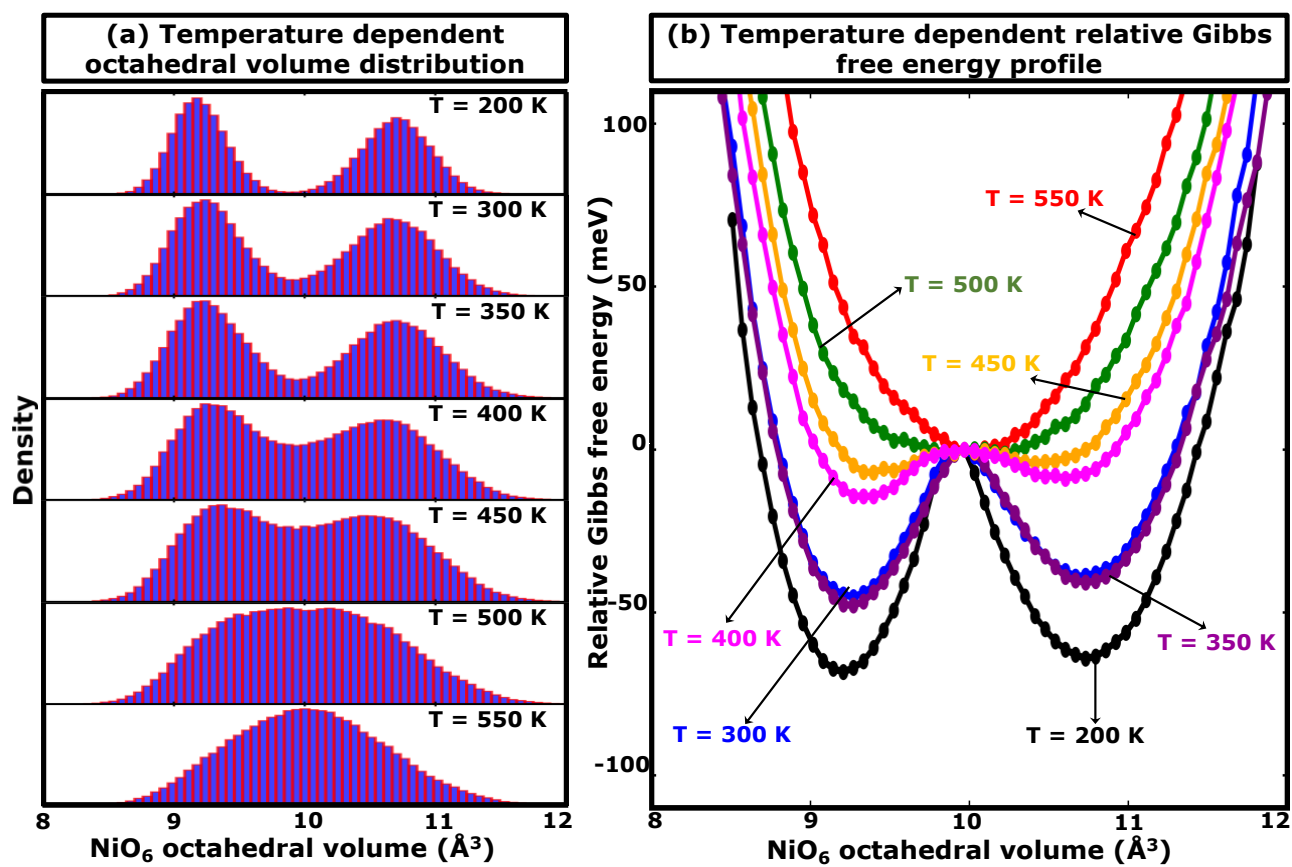

**Figure 4. Correlation of distribution of microstate properties to Gibbs free energy profile.**
(a) Distribution of local octahedra volumes for different supercell sizes extracted over 20 ps after initial equilibration of 5 ps. (b) Relative Gibbs free energy profile computed based on the distribution of octahedra volumes. The results are presented based on the thermal equilibration of the monoclinic (β) PM phase (as starting configuration) modeled as a 160-atom supercell with a polymorphous spin distribution.

**Potential surface profile extracted from first-principles theory at finite temperature:** Knowing the distribution of local structural motifs, one may wonder if extracting the effective coordinate-dependent potential surface energy is possible. This is critical as it can potentially demonstrate the origin of the potential phase transition and, more generally, explain the role of temperature in the distribution of local motifs. We note that describing the presence of a double local environment leads to a double-well free energy potential.[40] The relationship between statistical physics and thermodynamics allows for the computation of macroscopic properties like the Gibbs free energy ($\Delta G$) from the microscopic behavior of a system. AIMD simulations provide a series of "snapshots" representing different microstates, offering a statistical framework for understanding these thermodynamic quantities. Since different snapshots can be considered as non-interactive microstates of the systems, it is possible to apply statistical physics to extract energy profiles, specifically the Gibbs free energy can be expressed as $\Delta G(q) \approx -k_B T \ln \frac{p(q)}{p^{max}(q)}$, where $p(q)$ is a probability distribution of microstates and $p^{max}(q)$ corresponds to the highest probability among all microstates. Assuming that each microstate of the system can be characterized by the distribution of octahedra volumes extracted over time (we note, however, that such an assumption requires converged distribution of local motifs with time (Fig. 4a)), we calculated the relative Gibbs free energy profile shown in Fig. 4b. Based on the computed results, one can see the following details. At the low temperature, there is a distinct double well picture showing that the system tends to structurally disproportionation (i.e., form small and large $NiO_6$ octahedra), which is consistent with that described by Ghosez et al.[40] It should be noted here that imperfection in the potential surface energy profile is mainly caused by the rather short sampling time (about 15 ps for each temperature after the initial equilibration) and may completely converge with increasing time. As temperature increases to about 500-550 K, there is suppression of the double well distribution, resulting in a single well Gibbs free energy profile. These results thus demonstrate how the temperature can significantly modify the potential energy profile going from double well to single well, and indeed, it is the temperature responsible for the suppression of spontaneous energy lowering disproportionation. The important part of these results is that the depth of the double well energy profile slowly reduces with increasing temperature. This picture thus demonstrates that the phase transition in $YNiO_3$ is indeed similar to that observed in some other materials that are considered to be correlated.[42] Given the temperature-dependent distribution of local motifs and comprehension of the Gibbs free energy profile, it is imperative to understand the novel insights derived from this data. First, this study illustrates the power of the spin polymorphous model in accurately representing the transition from the monoclinic to orthorhombic phase in paramagnetic $YNiO_3$. Despite observing anticipated patterns at reduced temperatures, a crucial observation is that certain defining features dissipate earlier (e.g., already at low temperatures, experimentally observed s-XRD patterns[25] show only the barely distinguishable peak splits ($2\theta \sim 22.6°$) used to identify the presence of monoclinic phase) than the suppression of the disproportionation temperature. Thus, concentrating on only one aspect might not provide an accurate representation of phase transition temperatures, especially if they are defined based on the suppression of structural disproportionation temperatures. This observation aligns with experimental results, reinforcing the effectiveness of the spin polymorphous model in considering temperature-dependent factors. Second, we also observe that while heating, as the system approaches the transition temperature (we note, however, that the exact transition temperature may indeed depend on some computational factors such as time of molecular dynamic equilibration, the density of k-points, etc.), some interesting physics can be observed. For instance, in Fig. 2, the time-resolved distribution of octahedral motifs fluctuates between single and double-peak octahedral volume distributions. This opens up interesting avenues in the physics of quantum materials close to the transition temperature. These observations demonstrate the complexity of phase transitions near the phase transition temperature. We note that such complex symmetry breaking cannot be easily detected with standard techniques like s-XRD. This further emphasizes the importance of using methods that can distinguish local symmetry breaking, such as pair distribution analysis.

**Conclusions:** In summary, using a combination of density functional theory and ab initio molecular dynamics within the polymorphous framework, we provide a fundamental understanding of the distribution of local spin and structural motifs in paramagnetic $YNiO_3$ phases, showing the existence of two types of octahedra (large and small) that play a key role in the insulating nature of this compound. Specifically, we focus on the temperature effect on the distribution of local motifs and provide comprehensive insights into the thermally induced transformation of these motifs and its correlation to the potential surface profile. For instance, we observe that the energy potential profile changes from a double-well at low temperatures to a single-well profile at around 500-550 K. These results thus show how temperature modifies the potential energy profile, reflecting the temperature-induced phase transition. Taking into account that the properties of the given system should be calculated as some superposition of properties of local motifs (and not as properties of global average structure), we thus highlight the need to account for the temperature-induced distribution of local motifs in describing properties of correlated materials. Moreover, we also reaffirmed the limitations of relying solely on X-ray diffraction peaks to determine phase transitions. Our findings hint that considering the degree of structural disproportionation could offer a more precise indication of a phase transition. However, despite the promising results, the study recognizes that the simulations and models utilized have limitations and may not precisely capture the intricate behavior of these systems.

**Acknowledgment:** The authors thank the "ENSEMBLE[3] - Centre of Excellence for nanophotonics, advanced materials and novel crystal growth-based technologies" project (GA No. MAB/2020/14) carried out within the International Research Agendas programme of the Foundation for Polish Science co-financed by the European Union under the European Regional Development Fund and the European Union's Horizon 2020 research and innovation programme Teaming for Excellence (GA. No. 857543) for support of this work. We gratefully acknowledge Poland's high-performance computing infrastructure PLGrid (HPC Centers: ACK Cyfronet AGH) for providing computer facilities and support within computational grant no. PLG/2023/016228. The authors acknowledge fruitful discussions with Prof. Andriy Gusak.

**AUTHOR DECLARATIONS**
**Conflict of Interest**
The authors have no conflicts to disclose.

## Author Contributions

**Himanshu Joshi:** Data curation (lead); Formal analysis (equal); Investigation (lead); Methodology (equal); Validation (equal); Writing – original draft (supporting).

**Mateusz Wlazło:** Data curation (supporting); Formal analysis (equal); Investigation (supporting); Methodology (equal); Validation (supporting); Writing – original draft (supporting).

**Harshan Reddy Gopidi:** Data curation (supporting); Formal analysis (supporting); Investigation (supporting); Methodology (equal); Validation (supporting); Writing – original draft (supporting).

**Oleksandr I. Malyi:** Data curation (supporting); Formal analysis (equal); Investigation (supporting); Methodology (equal); Validation (supporting); Writing – original draft (lead), Project administration (lead); Supervision (lead).

## DATA AVAILABILITY

The data that support the findings of this study are available from the corresponding author upon reasonable request.